\documentclass[conference]{IEEEtran}
\IEEEoverridecommandlockouts
\usepackage{cite}
\usepackage{amsmath,amssymb,amsfonts}
\usepackage{algorithmic}
\usepackage{graphicx}
\usepackage{textcomp}
\usepackage{xcolor}
\def\BibTeX{{\rm B\kern-.05em{\sc i\kern-.025em b}\kern-.08em
    T\kern-.1667em\lower.7ex\hbox{E}\kern-.125emX}}

\usepackage[nolist]{acronym}

\usepackage{tabularx}
\usepackage{soul,color}
\usepackage{subcaption}
\usepackage{authblk}

\usepackage[normalem]{ulem}
\usepackage{adjustbox}

\usepackage{enumitem}
\usepackage{booktabs}

\usepackage{balance}

\begin{document}

\title{Spreading Factor assisted LoRa Localization with Deep Reinforcement Learning}

\author[1]{Yaya Etiabi}
\author[1]{Mohammed Jouhari}
\author[2]{Andreas Burg}
\author[1]{El Mehdi Amhoud}
\affil[1]{School of Computer Science, Mohammed VI Polytechnic University, Benguerir, Morocco }
\affil[2]{Telecommunication Circuits Laboratory, École Polytechnique Fédérale de Lausanne, Lausanne, Switzerland \authorcr Email: {\{yaya.etiabi, mohammed.jouhari, elmehdi.amhoud\}@um6p.ma, andreas.burg@epfl.ch}}

\maketitle

\begin{abstract}
Most of the developed localization solutions rely on RSSI fingerprinting. However, in the LoRa networks, due to
the spreading factor (SF) in the network setting, traditional fingerprinting may lack representativeness of the radio map, leading to inaccurate position estimates. As such, in this work, we propose a novel LoRa RSSI fingerprinting approach that takes into account the SF. The performance evaluation shows
the prominence of our proposed approach since we achieved an improvement in localization accuracy by up to 6.67\% compared to the state-of-the-art methods. The evaluation has been done using a fully connected deep neural network (DNN) set as the baseline. To further improve the localization accuracy, we propose a deep reinforcement learning model that captures the ever-growing complexity of LoRa networks and copes with their scalability. The obtained results show an improvement of 48.10\% in the localization accuracy compared to the baseline DNN model.
\end{abstract}


\section{Introduction}
The Internet of Things (IoT) is continuously reshaping the way we interact with our environment. This is done thanks to the widespread of cutting-edge sensing devices deployment at an unprecedented pace \cite{etiabi2020}, and the development of low-cost and long-range communication technologies such as NB-IoT, SigFox, LTE-M, and LoRaWAN. The latter is the most widely used IoT communication technology and seems very promising for IoT applications involving limited resources, regarding its low cost, long-range communication capability, and its ease of deployment, as shown in \cite{Jouhari2022ASO}.
With the increasing need for accurate positioning systems in these ever-growing IoT networks,
GPS-based solutions
fail to
support the location-based services either due to environmental challenges or due to the cost of their deployment.
As a result, many research works have attempted to provide LoRa-powered localization systems by exploiting the LoRa networks properties such as fingerprinting, time of arrival (ToA), angle of arrival (AoA), and time difference of arrival (TDoA) 
\cite{9860962}.
More recently, several machine learning (ML) algorithms have been proposed to enhance the localization accuracy and to reduce the online prediction complexity involved in the traditional methods \cite{Almeida2021BlindTL,Etiabi2023FederatedLB,MLSF}. These ML techniques are mostly based on the fingerprints prerecorded in the network.
Nonetheless, unlike WiFi \cite{etiabi2022federated}, in the LoRa communication, such RSSI fingerprints may lack representativeness of the network radio map since the RSSI fingerprinting of the covered area is affected by the spreading factor (SF) which is a LoRa key parameter \cite{tapparel2021enhancing}. Indeed
the SF, an important metric in LoRaWAN communication, controls the number of chirps (data carriers) transmitted each second. With a higher SF, there are fewer chirps per second, and hence less data is processed per second. When transmitting the same amount of data with a greater SF, more airtime is required. 
To achieve more airtime, the LoRa node operates and runs for longer periods of time, costing more energy \cite{jouhari2023deep}. As a result, SF assignment is a performance-versus-energy-consumption trade-off that is critical for LoRa-based IoT applications.
The network determines the SF (graded 7-12) based on the environmental conditions between the end device and the gateway in order to fully use the benefits of LoRa technology and optimize its performance \cite{Azizi2022MIXMABRL}. 
Since the communication range is directly affected by the allocated SF, RSSI fingerprints are prone to high instability, especially in the LoRa networks implementing the adaptive data rate mechanism.

In this work, we propose a new localization framework that incorporates the SF in the construction of the network radio map.
Indeed, the objective of our study is to design an accurate and reliable LoRa localization
system in constrained and GPS-challenged outdoor environments. 
The motivation for using SF to build the radio map of the area of interest resides in the fact that different SFs lead to different RSSI fingerprints at a given location.
Thus, taking into consideration the SF will provide a fully comprehensive radio map that is expected to present better localization performance. Moreover, RSSI fingerprints collection may lead to complex datasets which are also impacted by the wireless channel conditions. The previous facts can mislead the deep neural network (DNN) which results in performance degradation of the overall localization system.
To overcome the latter issue, we propose a new formulation of the localization problem based on a deep reinforcement learning (DRL) approach which is expected to undertake the complex nature of the IoT environment and achieve better localization accuracy.
The contribution of our work is as follows:
\begin{itemize}
    \item First of all, we present a new radio map construction method for  the LoRa network.  Indeed, we investigate the construction of a more comprehensive radio fingerprints database with both the received signal levels and the different SFs used during the transmission.
    \item Afterwards, we show the prominence of using the SFs as an additional feature in the fingerprinting by assessing the impact of the SFs on the localization system accuracy.
    \item Finally, we tackle the localization problem in LoRa networks with a DRL approach using the RSSI along with the SF. Indeed, we propose the use of the DRL approach to capture the environmental complexity, the scalability issues in IoT networks, and above all, the localization system accuracy improvement.
\end{itemize}
The remainder of this paper is organized as follows. Section~\ref{sec:system-model} describes the SF-aware RSSI fingerprinting while Section~\ref{sec:problem-formulation} depicts our problem formulation. Section~\ref{sec:drl-localization}  introduces our DRL-based localization approach. In Section~\ref{sec:performance-eval}, we present the performance evaluation of our proposed method. Finally, in Section~\ref{sec:conclusion}, we conclude and set forth our perspectives.

\section{Spreading factor-aware RSSI fingerprinting}
\label{sec:system-model}
\begin{figure}[!t]
    \centering
    \includegraphics[scale=0.4]{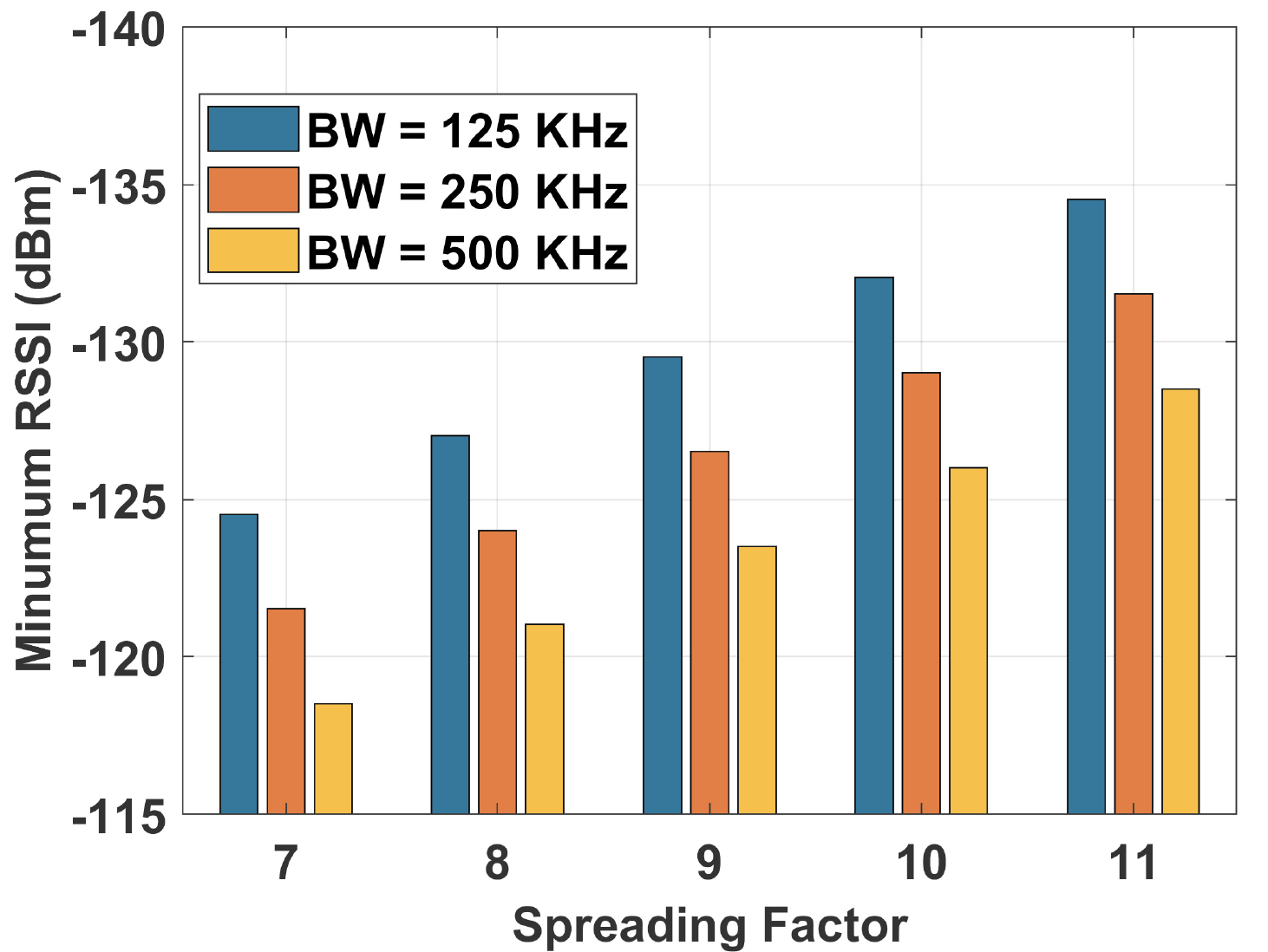}
    \caption{Minimum RSSI vs SF for different bandwidths}
    \label{fig:sf_rssi}
    \vspace{-0.1in}
\end{figure}

We consider a LoRa network covering an area of dimension $L\times W$, containing $M$ LoRa gateways distributed across the network. In LoRa as well as in other wireless networks, the RSSI is strongly related to the path loss experienced in the studied environment.
Basically,
using the log-distance path loss model as in \cite{Njima2021ConvolutionalNN},
the RSSI value captured at the $m^{th}$ gateway from the $n^{th}$ reference position at time $t$ can be expressed as:
\begin{equation}
  \begin{aligned}
RSSI_{n,t}^m &= P_{T_x} - \left[20\log _{10}\left(\frac{4\pi d_0}{c}\right)+20\log _{10}\left(f\right)+ \right. \\
&\left. 10\beta \log _{10}\left(\frac {d_n^m}{d_{0}}\right)+X_{\sigma,{n,m}}^t\right]\\
\end{aligned},
\label{equ:rssi}
\end{equation}
where $P_{T_x}$, $d_0$, $c$, $f$, $\beta$, $X_\sigma$ are respectively the transmitted power, the reference distance, the speed of light, the frequency of the signal, the path loss exponent, and the shadowing.
However, although LoRa has good sensitivity, depending on the channel conditions and the transmission range, poor signal reception can occur which may lead to missing RSSI recordings.
The recording of RSSI with respect to the receiver sensitivity is described by the following equation :
\begin{equation}
RSSI_{n,t}^m = \left\{\begin{matrix}
RSSI_{n,t}^m &\text{ if } RSSI_{n,t}^m>S_{n,t}^m\\ 
 NaN& \text{Otherwise}
\end{matrix}\right.  ,
\label{equ:rssi_rec}
\end{equation}
where \mbox{$S_{n,t}^m=-174+10 \log _{10} \text{BW} + \text{NF} +\text{SNR}_{n,t}^m$}
define
the sensitivities which depend on the received $\text{SNR}$ thresholds which in turn are related to the spreading factors $SF$.
%
Note that
$\mathrm{BW}$ is the bandwidth in $\mathrm{KHz}$,
$\mathrm{NF}$ the noise figure of the receiver in $\mathrm{dB}$, and
SNR$_{n,t}^m$ the signal-to-noise ratios.
Thus, as depicted in Fig.~\ref{fig:sf_rssi}, the minimum recorded RSSI value depends on the spreading factor allocated to the node, given a specific bandwidth. Overall, the higher the SF, the higher the sensitivity, and by extent, the longer the communication range and the better the coverage. 
As a consequence of the dynamic allocation of the SF, the constructed RSSI database may contain missing values (Eq.~\ref{equ:rssi_rec}) that need to be handled.
To do so, we consider the RSSI thresholds for different SF values so that a missing value is replaced by the RSSI threshold of the SF allocated during the communication round. This is indeed described by:
\begin{equation}
RSSI_{n,t}^m = \left\{\begin{matrix}
S_{n,t}^m(SF) &\text{ if } RSSI_{n,t}^m=NaN\\ 
 RSSI_{n,t}^m & \text{Otherwise}
\end{matrix}\right. .
\end{equation}
Afterwards, for each recording, the SF is added to the RSSI vector as an additional radio fingerprint.

\section{Problem Formulation}
\label{sec:problem-formulation}
We consider the localization system as a reinforcement learning (RL) agent whose goal is to accurately localize IoT devices in the LoRa network by interacting with the wireless environment. Our considered environment is defined by the RSSI signals and the geometry of the LoRa network, within which the agent shifts and changes a bounding square window ($\Omega$) through a sequence of actions and goes to the next state after performing a specific action under the current observation. When the targeted IoT device enters the environment and gets an RSSI signal, the localization agent is required to gradually locate it by enclosing it in a small enough window. The agent should identify how to move and rearrange the window at each phase of the localization process in order to efficiently localize the target in a few steps, as shown in Fig.~\ref{fig:dqrloc}.
Therefore, we formulate the localization problem as a dynamic Markov decision process characterized by the following components:
\begin{figure}[!t]
    \centering
    \includegraphics[scale=0.35]{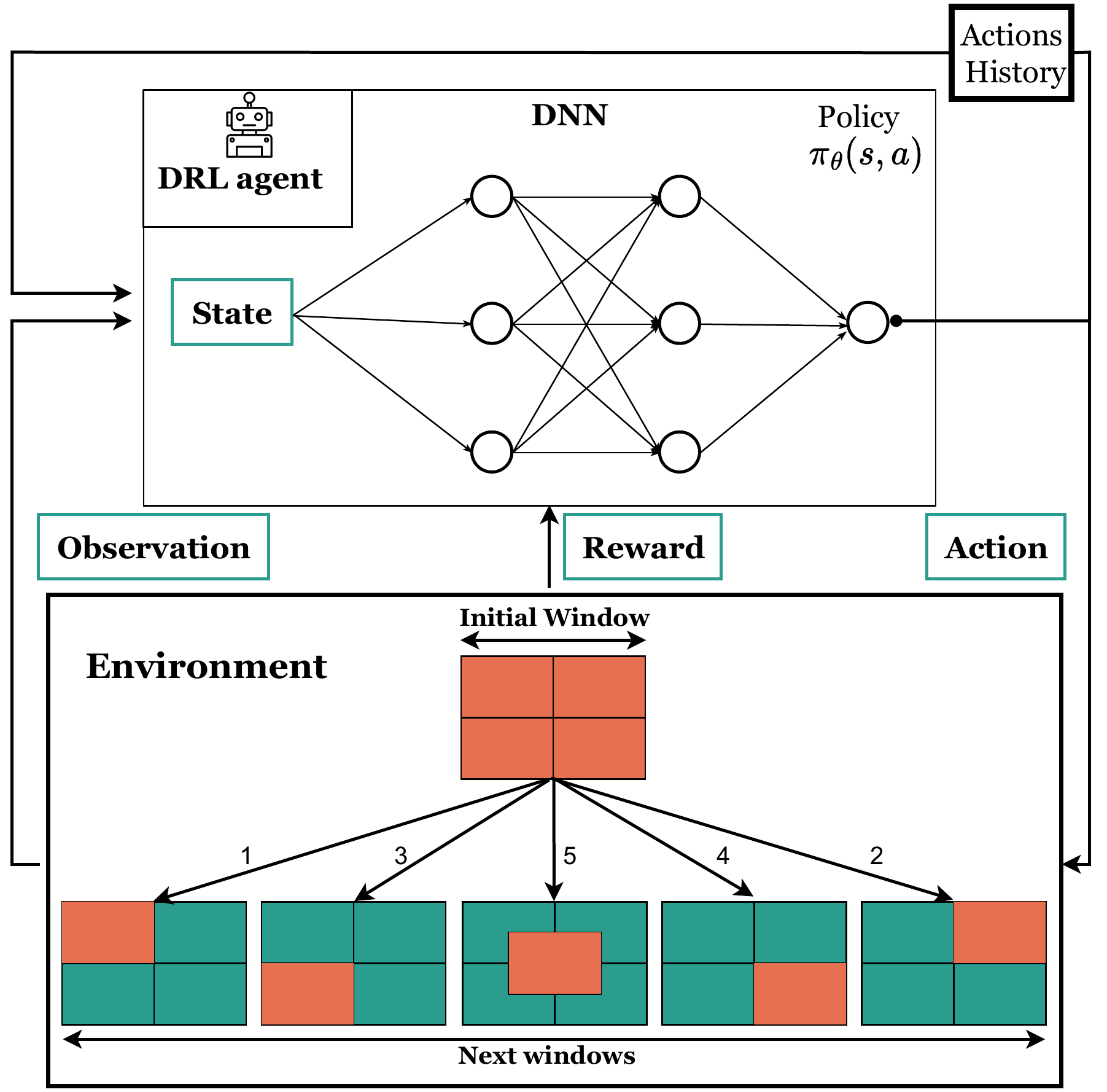}
    \caption{Deep reinforcement learning based  localization framework}
    \label{fig:dqrloc}
    \vspace{-0.15in}
\end{figure}
\subsubsection{State Space}
The state space of the localization problem encompasses the RSSI measurements, the SFs used during measurements, the search window $\Omega$ characterized by the center coordinates of the localization agent positions $\Theta = \left(x,\;  y\right)$, the window size $d$,  and the history vector $\mathcal{H}$ of actions taken so far. At given time step $t$, the state of the localization agent is defined by $s_t=\left [ RSSI_t, SF_t, \Omega_t, \mathcal{H}_t \right ]$ where $\Omega_t = \left [ \Theta_t, d_t \right ]$.

\subsubsection{Action Space}
The action space represents the set of actions the agent can perform when trying to locate a target. In our formulation, we distinguish 5 possible actions (1: up-left, 2: up-right, 3: down-left, 4: down-right, 5: center) as depicted in Fig.~\ref{fig:dqrloc} and defined as follows:
\begin{itemize}[align=left,leftmargin=*]
\item $a_t = 1$ : $
\Theta_{t+1}=(x_{t+1}, y_{t+1})=(x_{t}-\operatorname{d}_{t} / 2, y_{t}+\operatorname{d}_{t} / 2)$
\item $a_t = 2$ : $
\Theta_{t+1}=(x_{t+1}, y_{t+1})=(x_{t}+d_{t} / 2, y_{t}+d_{t} / 2)
$
\item $a_t = 3$ : $
\Theta_{t+1}=(x_{t+1}, y_{t+1})=(x_{t}-\operatorname{d}_{t} / 2, y_{t}-\operatorname{d}_{t} / 2)
$
\item $a_t = 4$ : $
\Theta_{t+1}= (x_{t+1}, y_{t+1})=(x_{t}+\operatorname{d}_{t} / 2, y_{t}-\operatorname{d}_{t} / 2)
$
\item $a_t = 5$ :  $
\Theta_{t+1}=(x_{t+1}, y_{t+1})=(x_{t}, y_{t})$ 
\end{itemize}
For each of the above actions, the search window is shrunk as follows: 
$\operatorname{d}_{t+1}=\alpha \times d_{t}$, with $\alpha=0.5$ in our formulation. 
As shown in Fig~\ref{fig:dqrloc}, the actions are encoded by their numerical values ranging from 1 to 5.
Thus, in order to avoid a misinterpretation of the history vector by the deep Q-network, the actions are one-hot encoded.


\subsubsection{Reward Function}
Our reward function is based on the normalized intersection of windows noted $\operatorname{IoW}\left(\Omega, \Omega^{*}\right) \in [0,\ 1]$ where $\Omega$ is the search window and $\Omega^*$ the target window. $\operatorname{IoW}\left(\Omega, \Omega^{*}\right)$ represents the overlapping area of the two windows normalized by the area of the search window and defined as:
\begin{equation}
\operatorname{IoW}\left(\Omega, \Omega^{*}\right)=\operatorname{area}\left(\Omega \cap \Omega^{*}\right) / \operatorname{area}(\Omega).
\end{equation}
The target window $\Omega^{*}$ is set to represent the smallest square enclosing the target. The search window is initialized at the beginning of the localization process as the smallest square covering the whole LoRa Network. Each time the RL agent takes an action, the search window $\Omega$ moves and shrinks, and the agent gets rewarded accordingly. This reward is defined as follows :
\begin{equation}
r_{a_{t}}\left(s_{t}, s_{t+1}\right)=\left\{\begin{aligned}
+\xi & \text {  if }  \operatorname{IoW}\left(\Omega^{s_{t+1}}, \Omega^{*}\right) &\in&[\Delta, 1] \\ 
+\phi & \text {  if }  \operatorname{IoW}\left(\Omega^{s_{t+1}}, \Omega^{*}\right) &\in &\\ 
&  \left[\operatorname{IoW}\left(\Omega^{s_{t}}, \Omega^{*}\right), \Delta\right[ &\\ 
-\xi & \text { otherwise}
\end{aligned}\right.,
\end{equation}
where $\Delta$ is the overlapping threshold of the search and the target windows, $\xi$ is the stopping reward, and $\phi$ is the reward for each good action that takes the localization agent closer to the target. In our simulations, these parameters have been set to 0.5, 10, and 1 respectively.

\section{DRL-based LoRa localization}
\label{sec:drl-localization}
Basically, the deep Q-network (DQN) was developed by enhancing the classic RL algorithm called Q-Learning with DNN and the technique known as experience replay \cite{dqn}.
Indeed, Q-learning
is based on the notion of a Q-function most often represented by the so-called Q-table that maps the expected return or discounted sum of rewards $Q^{\pi}(s, a)$ obtained from state $s$, by taking action $a$ and then following a policy $\pi$.
Then, the optimal Q-function $Q^{*}(s, a)$ obeys the Bellman optimality equation given by :
\begin{equation}
Q^{*}(s, a)=\mathbb{E}\left[r+\gamma \max _{a^{\prime}} Q^{*}\left(s^{\prime}, a^{\prime}\right)\right],
\label{equ:bellman}
\end{equation}
with $\gamma \in \left [0,1\right]$ is the discount factor which essentially determines how much the RL agent cares about rewards in the distant future relative to those in the immediate future.
Q-Learning intends to  estimate the Q-function by using the Bellman Eq.~\eqref{equ:bellman} as an iterative update formulated as :
\begin{equation}
Q_{i+1}(s, a) \leftarrow \mathbb{E}\left[r+\gamma \max _{a^{\prime}} Q_{i}\left(s^{\prime}, a^{\prime}\right)\right].
\end{equation}
Such iterative formulation converges to the optimal Q-function $Q_{i} \rightarrow Q^{*}$ as $i \rightarrow \infty$ as shown in \cite{atari}.
However, in localization problems, it is not possible to represent the Q-function as a table with values for each combination given the nature of the localization task where the state space is continuous.
Instead, we define a neural network-based approximation function with parameter $\theta$, known as the DQN such that $Q(s, a; \theta) \approx Q^{*}(s, a)$.
To do so, the  DQN is trained by iteratively adjusting the parameter $\theta_i$ to reduce the loss function given by :
\begin{equation}
L_{i}\left(\theta_{i}\right)=\mathbb{E}_{s, a, r, s^{\prime} \sim \rho(.)}\left[\left(y_{i}-Q\left(s, a ; \theta_{i}\right)\right)^{2}\right],
\label{equ:loss}
\end{equation}
$\text { where } y_{i}=r+\gamma \max _{a^{\prime}} Q\left(s^{\prime}, a^{\prime} ; \theta_{i-1}\right)$ and $\rho$ the distribution over transitions $\left\{s, a, r, s^{\prime}\right\}$ collected from the environment.

During the training, $\rho$ follows the $\epsilon$-greedy policy which chooses between a random action (exploration) with probability $\epsilon$ and a greedy action (exploitation) with probability $1-\epsilon$ to ensure good coverage of the state-action space. Moreover, at each training iteration, we decay the $\epsilon$ value so that the exploitation probability increases as the agent is learning. In standard Q-learning, the loss function is computed using only the last transition $\{s, a, r, s^{\prime}\}$. In our training, we use the \textit{Experience Replay}, a technique introduced in \cite{atari} to make the network updates more stable. Therefore, at each time step, the transition vector $\{s, a, r, s^{\prime}\}$ is stored in a circular buffer named the\textit{ replay buffer}. Then, we use  mini-batches from samples in the replay buffer to update the Q-network.

\section{Performance  Evaluation}
\label{sec:performance-eval}

\subsection{Baseline: LoRa Localization using RSSI and SF association fingerprints}
In this subsection, we investigate the prominence of using the SFs as important elements in addition to the RSSI fingerprints before feeding them to our DRL-based method.
To do so, we design a DNN architecture whose configuration is shown in TABLE~\ref{tab:dnn-config}. The training of the model can be seen as a minimization problem considering the loss function below:
\begin{equation}
\min _{\boldsymbol{W}} \frac{1}{N} \sum_{n=1}^{N} f\left(\boldsymbol{W}, \right[\boldsymbol{RSSI}_{{n}}\ SF_n\left], \boldsymbol{C}_{{n}}\right),
\end{equation}
where $\boldsymbol{C}_{{n}}$ is the $n^{th}$ reference position coordinates corresponding to the captured $\boldsymbol{RSSI}_{{n}}$ at that location, $f(\cdot)$ is the approximation error function, and ${\boldsymbol{W}}$ represents the weights or parameters of the DNN model.

\begin{figure}[!t]
    \centering
    \includegraphics[scale=0.7]{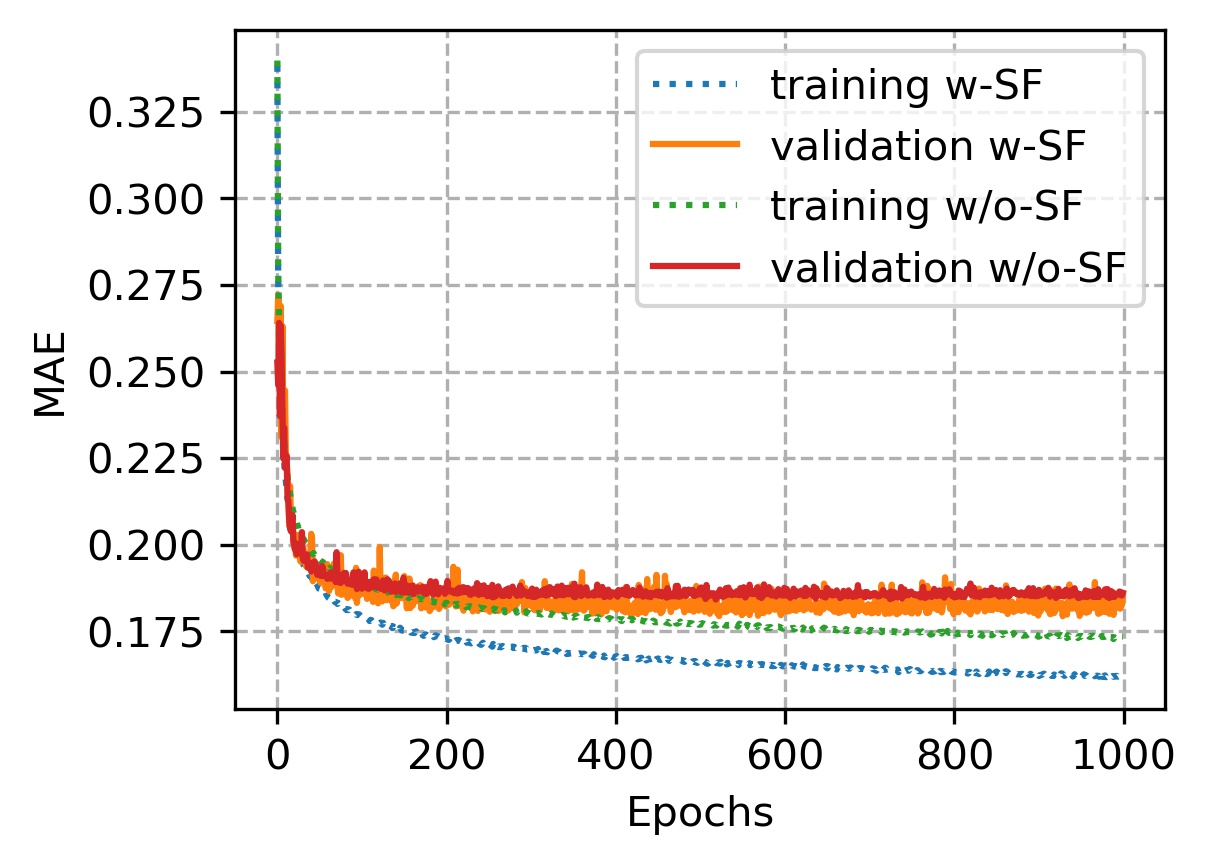}
    \caption{Mean absolute error of the baseline DNN localization with and without SF using experimental data \cite{urbanlora}}
    \label{fig:maeloss}
    \vspace{-0.15in}
\end{figure}

\begin{table}[t]
\vspace{4mm}
\caption{Parameters of the DNN model}
\centering
\resizebox{0.4\textwidth}{!}{\begin{tabular}{ m{4em}m{11em}m{11em} }
\toprule
 \textbf{Parameter} & \textbf{Description} & \textbf{Value}  \\
\bottomrule
 Optimizer & DNN Model Optimizer & Adam   \\

 $\mu$ & Learning rate & 0.0005    \\

$\beta_1, \beta_2$  & Exponential decay rates & $0.1, 0.99$  \\

$N_i$ & Input layers units & 10\\  

$N_h$ & Hidden layer units & $512\times256\times128\times64\times32$\\  

$D_1, D_2,D_3$ & Dropout layers & $0.3, 0.2, 0.1$\\  

$N_o$ & Output layer units & 2\\  

$\sigma_h(\cdot)$ & Hidden Activation function & ReLu\\  

$\sigma_o(\cdot)$ & Output Activation function & Linear\\  

$b$ & Batch size & 512\\  
\bottomrule
\end{tabular}}
\label{tab:dnn-config}
\end{table}
The optimal weights ${\boldsymbol{W}}^*$ are obtained through the minimization of the mean absolute error (MAE) loss function associated with the DNN model.
We use the LoRaWAN dataset published by Aernouts et al. in  \cite{urbanlora}, which was gathered over a 52 $km^2$ area in Antwerp, Belgium, by attaching LoRa modules to postal service vehicles and sending a message every minute to 68 LoRaWAN gateways. The experiment has been conducted for three months, resulting in 123,528 RSSI samples with the associated SFs used during the transmissions.
We set the input of the DNN model described in TABLE~\ref{tab:dnn-config} to $N_i=69$ (68 RSSI values from the 68 gateways + the transmission SF). The results are presented in Fig.~\ref{fig:maeloss} which reveals that with the SF as an additional feature input to the model, the convergence speeds up. 
The accuracy is improved by up to 6.67\% as shown in TABLE~\ref{tab:mde} where the evaluation is done using the mean distance error metric.
\begin{table}[t]
\vspace{4mm}
\caption{DNN localization error (m) with and without SF }
\centering
 \resizebox{0.4\textwidth}{!}{   \begin{tabular}{ c c c c  }
\toprule
 \textbf{Model} & \textbf{Training} & \textbf{Validation} & \textbf{Baseline experiment \cite{urbanlora}}  \\
\bottomrule
 w/o-SF &  319.84 & 365.03  & 398.40  \\
 w-SF & \textbf{298.51} & \textbf{359.71} & - \\
\bottomrule
\end{tabular}}
\label{tab:mde}
\vspace{-0.2in}
\end{table}

\subsection{Performance evaluation of the DRL system}
In the previous section, we showed the prominence of exploiting the SF as an additional feature for the construction of the LoRa network radio fingerprints. In this section, the improved fingerprint data with SF is used to assess the performance of our proposed DRL-based localization. 

\begin{figure*}[t]
\centering
\begin{subfigure}{.24\textwidth}
    \centering
    \includegraphics[width=1.05\linewidth]{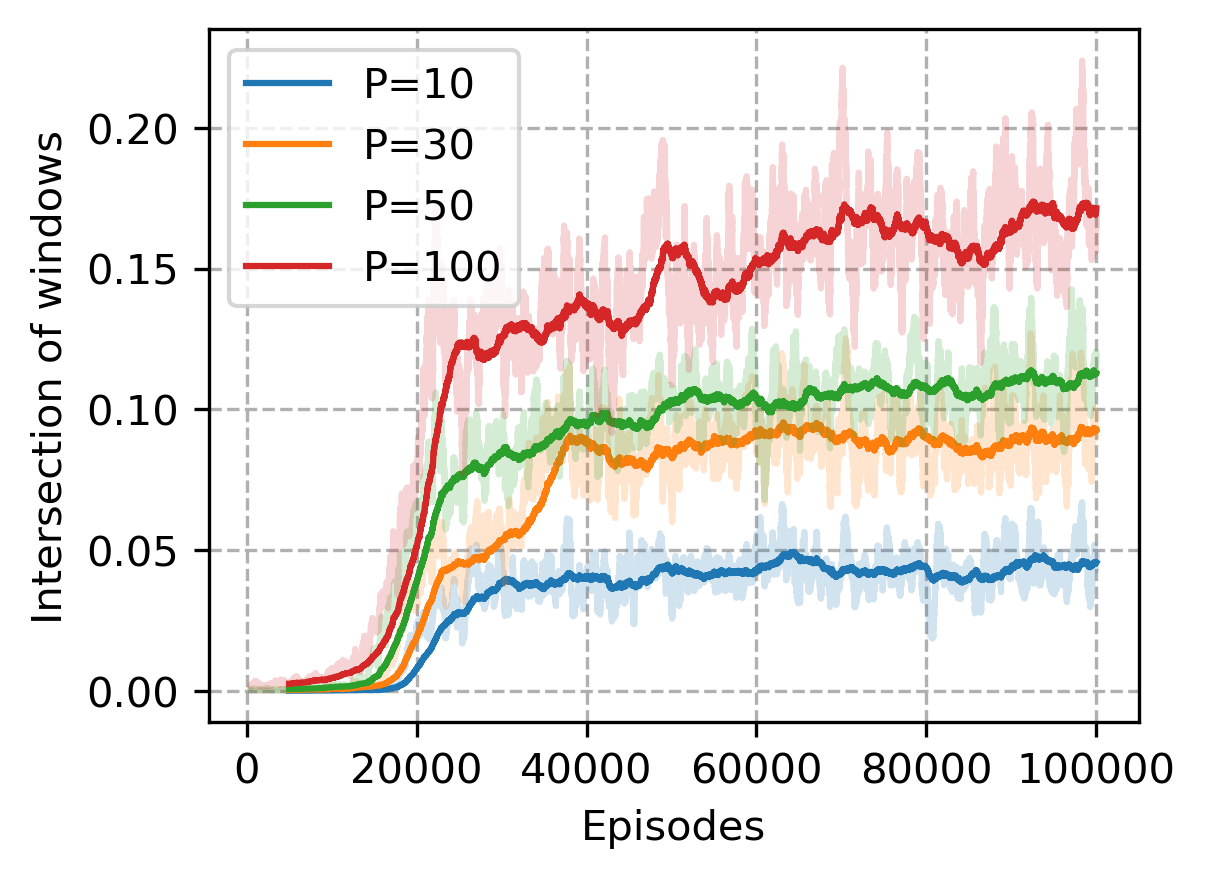}  
    \caption{}
    \label{fig:dqn-iow}
\end{subfigure}
\begin{subfigure}{.24\textwidth}
    \centering
    \includegraphics[width=1.05\linewidth]{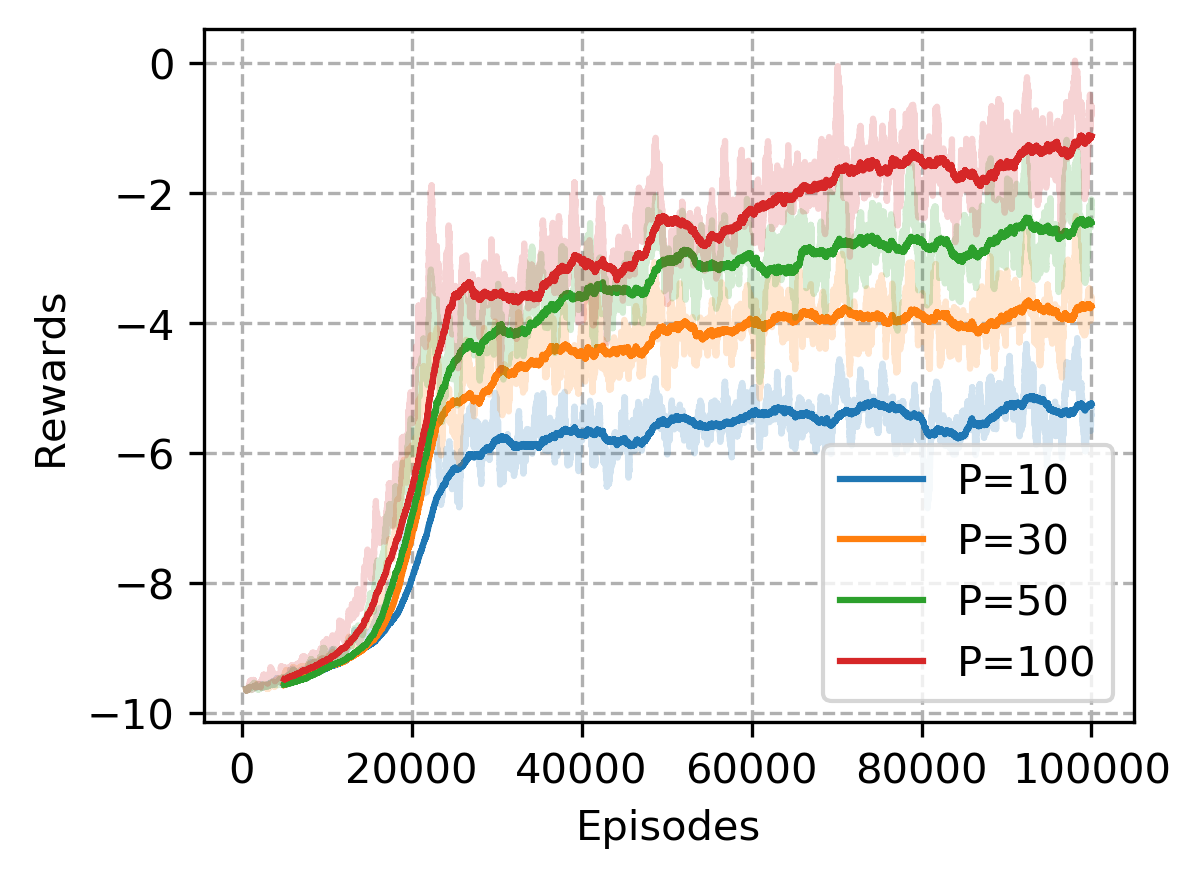}  
    \caption{}
    \label{fig:dqn-rewards}
\end{subfigure}
\begin{subfigure}{.24\textwidth}
    \centering
    \includegraphics[width=1.05\linewidth]{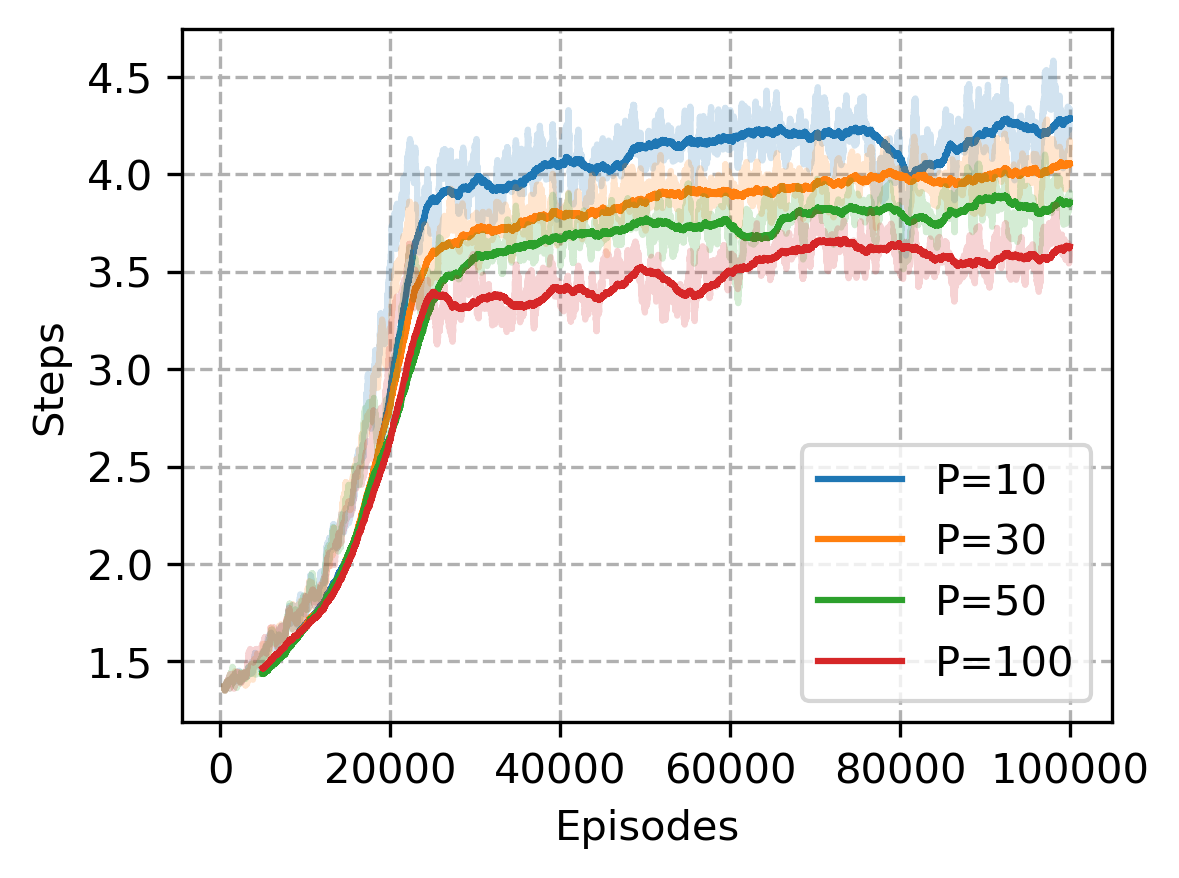}  
    \caption{}
    \label{fig:dqn-steps}
\end{subfigure}
\begin{subfigure}{.24\textwidth}
    \centering
    \includegraphics[width=1.05\linewidth]{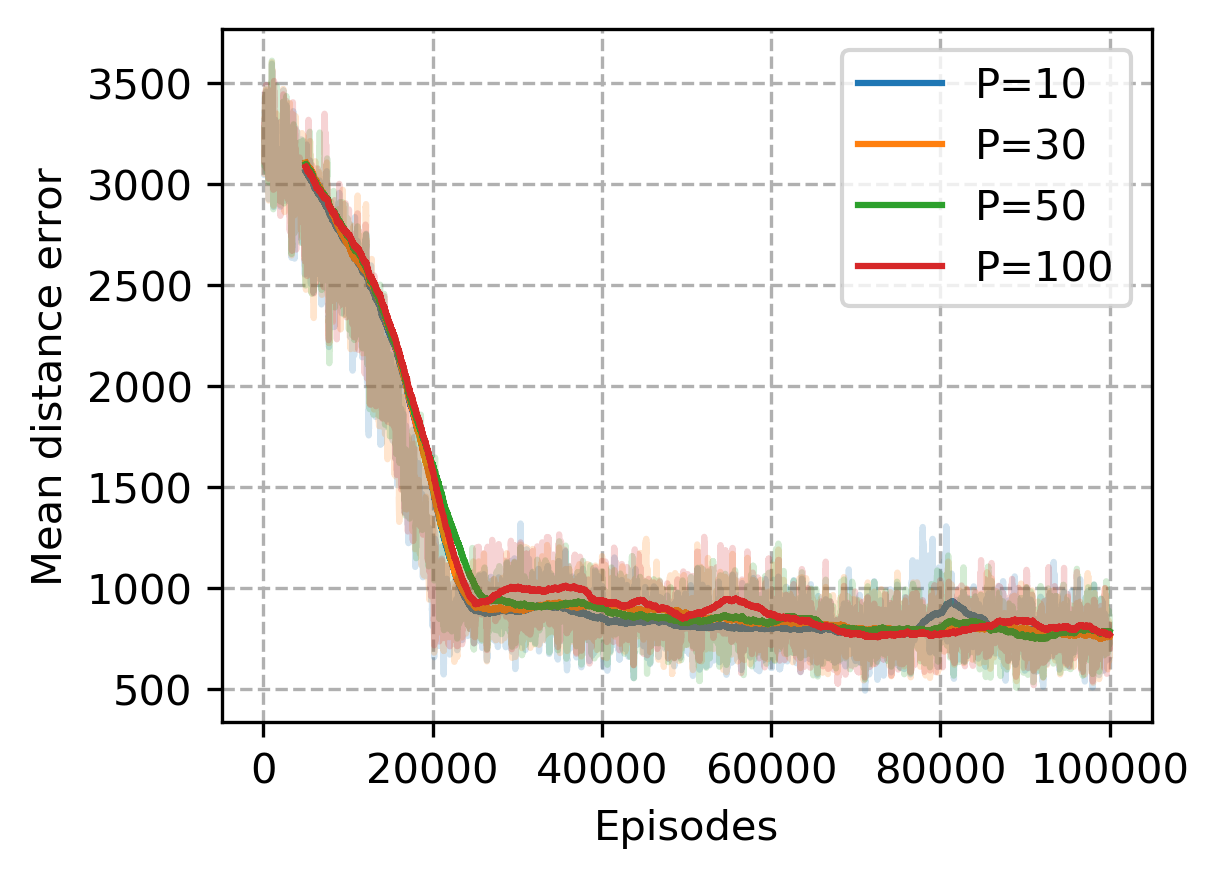}  
    \caption{}
    \label{fig:dqn-mde}
\end{subfigure}
\caption{Training performance of the DQN network for SF-assisted localization using experimental data\cite{urbanlora}: (a) Intersection of windows per episode (b) Reward per episode (c) Number of steps per episode (d) Mean distance error per episode}
\label{fig:dqn-preformance}
\vspace{-0.1in}
\end{figure*}


\begin{figure}[t]
    \centering
    \includegraphics[scale=0.8]{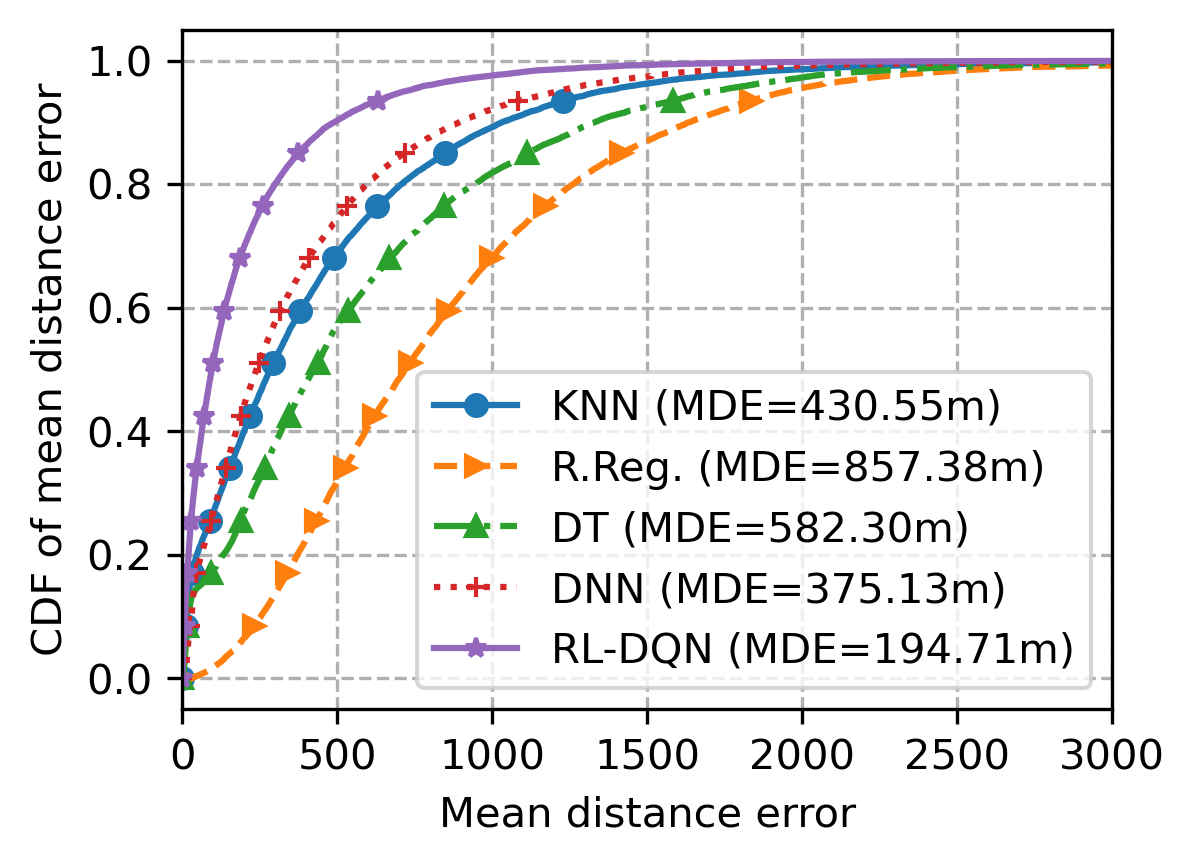}  
\caption{Performance evaluation and benchmarking of the DQN network  for SF-assisted localization} 
\label{fig:dqn-benchmark-cdf}
\vspace{-0.15in}
\end{figure}

To do so, since our DRL-based localization is a top-down process, we start by defining an initial search window $\Omega_0$ representing the smallest square that contains the whole LoRa network, and the target window $\Omega^*$ which determines how accurate we want our model to be. 
Using the experimental dataset of \cite{urbanlora}, each sample corresponds to an episode for our localization agent and its target window is defined by the corresponding label as  $\Omega^* = ({\Theta}^*=(x^*,y^*), d^*)$, where $x^*,\;y^*$ represent respectively the longitude and the latitude of the combined RSSI and SF sample. $d^*$ is the half length of the squared window centered at ${\Theta}^*$ and represents the chosen precision $P$ for our system.
The initial search window is characterized by its center coordinates $\Theta_0$ and its half-length $d_0$ defined  respectively by
\mbox{${\Theta}_0 = \left( \frac{x_{min}+x_{max}}{2}, \frac{y_{min}+y_{max}}{2} \right)$} and \mbox{${d}_0 = \frac{max\left(x_{max}-x_{min},y_{max}-y_{min}\right)}{2} +d^*$}
where $d^* = P$ is set empirically. We configure the deep Q-network with $128\times128\times64$ hidden layers, a learning rate of $5\times10^{-4}$, a batch size of 512, and a replay memory of length 50000.
The remaining parameters are set as follows: the $\epsilon$-greedy policy $\epsilon=1$ (decaying at each training time step until $\epsilon_{min}=0.001$), the discount factor $\gamma=0.1$, and the actions history vector size $\left| \mathcal{H}\right|=10$.
For the experiment, we split the dataset into training and test sets resulting in 100,000 and 23,528 samples for each, respectively.
The experiment has been done using a customization of \textit{OpenAI Gym} environment and the \textit{Stable-Baselines3} RL library.

 We report the training performance of the DQN in Fig.~\ref{fig:dqn-preformance} using different sizes of the target windows which reflects the precision constraints of the localization system. Indeed, Fig.~\ref{fig:dqn-iow}, points out the improvement of the intersection of windows (IoW) over the training episodes. Note that the IoW is crucial for the episode termination. It is used as a signal to indicate whether an episode has to be ended. 
 We can observe that the IoW is much higher with a large target window size, i.e. the larger the target window, the higher the IoW.
 Also, the localization agent is rewarded based on the evaluation of its current IoW, hence the performance reported in Fig.~\ref{fig:dqn-rewards} regarding the agent rewards over training episodes. Therefore, the rewards over the training episodes follow the same dynamics as of the IoW.
 On the other hand, Fig.~\ref{fig:dqn-steps} depicts the dynamics of the number of searching steps the agent is required to perform in order to localize the target. It can be seen that, small target windows required more searching steps to be found. This remark bears the impact of the target window size on the localization accuracy depicted in Fig.~\ref{fig:dqn-mde}. Note that at each searching step, the size of the search window $d$ is divided by two. That is, at the end of each episode, the final window within which the target is estimated to follow a power-of-2 decay law with respect to the number of steps. Indeed, if it takes $n$ steps to localize the target, the final size of the search window is $d_n = d_0/2^{n}$.
Finally, we evaluated the trained DQN performance with $P=10m$ along with a benchmark analysis using the test data.
For the benchmark analysis, we consider various state-of-the-art localization methods including K-nearest neighbors (KNN) where we set K to 11, Ridge regression (R.Reg.), decision tree (DT) with a depth of 10, and the previous baseline DNN model described in TABLE~\ref{tab:dnn-config}.
In Fig.~\ref{fig:dqn-benchmark-cdf}, we plotted the cumulative distribution function of the localization errors of the different algorithms. We also highlighted therein the mean error of each. It can be seen that our proposed method (RL-DQN) outperforms the studied algorithms with an improvement in the localization accuracy by up to 48,10\% compared to the DNN which is the most accurate one among the benchmarked algorithms.
\vspace{-0.1in}
\section{Conclusion}
\label{sec:conclusion}
In this paper, we proposed a novel LoRa fingerprinting technique that uses a combination of RSSI recordings and the spreading factors
to construct a more comprehensive LoRa network radio map.
The evaluation findings 
demonstrate that our method can increase localization accuracy by up to 6.67\%  compared to state-of-the-art systems. 
Additionally, in response to the accuracy requirement in an increasingly complex IoT environment, we proposed a deep reinforcement learning-based localization system which showed very promising results with an improvement of 48,10\% in accuracy compared to the baseline DNN model.
To further improve our results, in addition to a more robust RSSI data preprocessing technique and a hyper-parameter tuning, a joint optimal SF allocation and high accuracy localization method will be investigated in future work.

\vspace{-0.05in}
\section*{Acknowledgment}
This work was sponsored by the Junior Faculty Development program under the UM6P-EPFL Excellence in Africa
Initiative
\vspace{-0.05in}
\balance
\bibliographystyle{IEEEtran}
\bibliography{main}

\begin{thebibliography}{10}
\providecommand{\url}[1]{#1}
\csname url@samestyle\endcsname
\providecommand{\newblock}{\relax}
\providecommand{\bibinfo}[2]{#2}
\providecommand{\BIBentrySTDinterwordspacing}{\spaceskip=0pt\relax}
\providecommand{\BIBentryALTinterwordstretchfactor}{4}
\providecommand{\BIBentryALTinterwordspacing}{\spaceskip=\fontdimen2\font plus
\BIBentryALTinterwordstretchfactor\fontdimen3\font minus
  \fontdimen4\font\relax}
\providecommand{\BIBforeignlanguage}[2]{{%
\expandafter\ifx\csname l@#1\endcsname\relax
\typeout{** WARNING: IEEEtran.bst: No hyphenation pattern has been}%
\typeout{** loaded for the language `#1'. Using the pattern for}%
\typeout{** the default language instead.}%
\else
\language=\csname l@#1\endcsname
\fi
#2}}
\providecommand{\BIBdecl}{\relax}
\BIBdecl

\bibitem{etiabi2020}
Y.~Etiabi, E.~M. Amhoud, and E.~Sabir, ``A distributed and collaborative
  localization algorithm for internet of things environments,'' in
  \emph{Proceedings of the 18th International Conference on Advances in Mobile
  Computing \& Multimedia}, 2020, pp. 114--118.

\bibitem{Jouhari2022ASO}
M.~Jouhari, E.-M. Amhoud, N.~Saeed, and M.-S. Alouini, ``A survey on scalable
  {LoRaWAN} for {Massive IoT}: Recent advances, potentials, and challenges,''
  \emph{ArXiv}, vol. abs/2202.11082, 2022.

\bibitem{9860962}
D.~Merhej, I.~Ahriz, S.~Garcia, and M.~Terré, ``{LoRa} based indoor
  localization,'' in \emph{IEEE 95th Vehicular Technology Conference:
  (VTC2022-Spring)}, 2022, pp. 1--5.

\bibitem{Almeida2021BlindTL}
I.~B.~F. de~Almeida, M.~Chafii, A.~Nimr, and G.~Fettweis, ``Blind transmitter
  localization in wireless sensor networks: A deep learning approach,''
  \emph{32nd Annual International Symposium on Personal, Indoor and Mobile
  Radio Communications (PIMRC)}, pp. 1241--1247, 2021.

\bibitem{Etiabi2023FederatedLB}
Y.~Etiabi, W.~Njima, and E.-M. Amhoud, ``Federated learning based hierarchical
  {3D} indoor localization,'' in \emph{{IEEE} Wireless Communications and
  Networking Conference ({WCNC})}, 2023, {A}ccepted.

\bibitem{MLSF}
C.~J. Bouras, A.~Gkamas, S.~A.~K. Salgado, and N.~Papachristos, ``A comparative
  study of machine learning models for spreading factor selection in {LoRa}
  networks,'' \emph{Int. J. Wirel. Networks Broadband Technol.}, vol.~10, pp.
  100--121, 2021.

\bibitem{etiabi2022federated}
Y.~Etiabi, M.~Chafii, and E.-M. Amhoud, ``Federated distillation based indoor
  localization for {IoT} networks,'' \emph{ArXiv}, vol. abs/2205.11440, 2022.

\bibitem{tapparel2021enhancing}
J.~Tapparel, M.~Xhonneux, D.~Bol, J.~Louveaux, and A.~Burg, ``Enhancing the
  reliability of dense {LoRaWAN} networks with multi-user receivers,''
  \emph{IEEE Open Journal of the Communications Society}, vol.~2, pp.
  2725--2738, 2021.

\bibitem{jouhari2023deep}
M.~Jouhari, K.~Ibrahimi, J.~B. Othman, and E.~M. Amhoud, ``Deep reinforcement
  learning-based energy efficiency optimization for flying {LoRa} gateways,''
  \emph{arXiv preprint arXiv:2302.05214}, 2023.

\bibitem{Azizi2022MIXMABRL}
F.~Azizi, B.~Teymuri, R.~Aslani, M.~Rasti, J.~Tolvanen, and P.~H.~J. Nardelli,
  ``{MIX-MAB}: Reinforcement learning-based resource allocation algorithm for
  {LoRaWAN},'' \emph{IEEE 95th Vehicular Technology Conference:
  (VTC2022-Spring)}, pp. 1--6, 2022.

\bibitem{Njima2021ConvolutionalNN}
W.~Njima, M.~Chafii, A.~Nimr, and G.~Fettweis, ``Convolutional neural networks
  based denoising for indoor localization,'' \emph{93rd Vehicular Technology
  Conference (VTC2021-Spring)}, pp. 1--6, 2021.

\bibitem{dqn}
V.~Mnih, K.~Kavukcuoglu, D.~Silver, and et~al., ``Human-level control through
  deep reinforcement learning,'' \emph{Nature}.

\bibitem{atari}
V.~Mnih, K.~Kavukcuoglu, D.~Silver, A.~Graves, I.~Antonoglou, D.~Wierstra, and
  M.~A. Riedmiller, ``Playing {Atari} with deep reinforcement learning,''
  \emph{ArXiv}, vol. abs/1312.5602, 2013.

\bibitem{urbanlora}
M.~Aernouts, R.~Berkvens, K.~Van~Vlaenderen, and M.~Weyn, ``{Sigfox} and
  {LoRaWAN} datasets for fingerprint localization in large urban and rural
  areas,'' \emph{MDPI Data}, vol.~3, no.~2, p.~13, 2018.

\end{thebibliography}

\end{document}